\documentclass{iaus}
\def\beg{\begin{eqnarray}}
\def\ende{\end{eqnarray}}
\def\gsim{\lower.4ex\hbox{$\;\buildrel >\over{\scriptstyle\sim}\;$}} 
\def\lsim{\lower.4ex\hbox{$\;\buildrel <\over{\scriptstyle\sim}\;$}}

\usepackage{graphicx}
\renewcommand{\vec}[1]{\mbox{\boldmath$#1$}}
\title[On radiation-zone dynamos] 
{On radiation-zone dynamos}

\author[G\"unther R\"udiger  \& Marcus Gellert \& Rainer Arlt]   
{G\"unther R\"udiger,
 Marcus Gellert \and Rainer Arlt}

\affiliation{Astrophysikalisches Institut Potsdam,  \\ An der Sternwarte 16,
D-14482 Potsdam, Germany \\ email: {\tt gruediger@aip.de, \ mgellert@aip.de, \ rarlt@aip.de} \\[\affilskip]}

\pubyear{2010}
\volume{274}  
\pagerange{}
\jname{Advances in Plasma Astrophysics}
\editors{A. Bonanno,  E. de Gouveia dal Pino \& A. Kosovichev, eds.}
\begin{document}

\maketitle

\begin{abstract}
It is shown that the magnetic current-driven (`kink-type') instability produces flow and field patterns with helicity and even with $\alpha$-effect but only if the magnetic background field possesses non-vanishing current helicity  $\bar{\vec{B}}\cdot {\rm curl} \bar{\vec{B}}$ by itself. Fields with positive large-scale current helicity lead to  negative small-scale kinetic  helicity. The resulting $\alpha$-effect is positive. These results are very   strict for cylindric setups without  $z$-dependence of the background fields. The sign rules also hold for the more complicated cases in spheres where the toroidal fields are the result of the  action of differential rotation (induced from fossil poloidal fields) at least for the case that the global rotation is switched off after the onset of the instability. 
\end{abstract}

\firstsection 
\section{Introduction}
Open questions in stellar physics led to the idea that a dynamo operates in the radiative cores of early-type stars (Spruit 2002). Even the helioseismologic observation of rigid rotation of the solar interior shows in this direction. The angular momentum transport by the large-scale magnetic field pattern (fossil field plus toroidal field induced by differential rotation) does {\em not} lead to a solid-body rotation  unless the viscosity of the plasma exceeds the molecular value by a few orders of magnitude (R\"udiger \& Kitchatinov 1996, Eggenberger et al. 2005). Other examples are given by the evolution of the fast rotating early-type stars which can only be understood if i) there is a basic transport of angular momentum outwards and ii) the radial mixing of chemicals remains  weak (Yoon et al. 2006, Brott et al. 2008). Hence, if a  (magnetic-induced) instability existed in the radiative stellar cores then the corresponding Schmidt number  ${\rm Sc}= \nu/D$ must be rather large. We have shown that the kink-type instability (or Tayler instability, TI) of toroidal magnetic fields forms a much-promising candidate for the instability. A Schmidt number larger than ten  results as the ratio of the effective viscosity and the diffusion coefficient  (R\"udiger et al. 2009).

The unstable modes of the TI are basically  nonaxisymmetric  driven by the energy of the electrical current which produces the toroidal field. Interestingly enough, there exists even an instability of a toroidal magnetic field which in the fluid is current-free. In this case the energy comes from a differential rotation which itself is stable but which is unstable under the influence of the (current-free) toroidal field. We have called this instability as Azimuthal MagnetoRotational Instability (AMRI) as -- like for the standard MRI (with axial fields) -- the field itself is current-free and does not exert forces. In opposition to the standard MRI  the AMRI is always  nonaxisymmetric and it is, therefore,  much more interesting for  the dynamo theory. For  complicated  radial profiles of the toroidal field we shall always have a mixture of TI and AMRI. Generally, the latter is more important for fast rotation ($\Omega > \Omega_{\rm A}$) and v.v.  Here the Alfv\'{e}n frequency $\Omega_{\rm A}$ for the toroidal field is used which derives from the Alfv\'{e}n velocity $v_{\rm A}= B_\phi/\sqrt{\mu_0 \rho}$ as the related frequency. Between two cylinders with different radii the toroidal field profile with $B_\phi = A R + B/R$ ($R$ radius) is free of dissipation. The `perfect' AMRI appears for $A=0$ while the `perfect' TI results for $B=0$.

One can compute the necessary electrical currents to excite both sorts of instabilities in a columnar Taylor-Couette experiment with gallium as fluid conductor. As we have shown the critical Hartmann numbers for self-excitation of axi- and nonaxisymmetric perturbation modes do not depend on the magnetic Prandtl number of the fluid which is as small as $10^{-3}$ for stellar plasma and  $10^{-5}$ for liquid sodium (R\"udiger \& Schultz 2010).

It is typical for the nonaxisymmetric TI and AMRI that always the two modes with $m=\pm 1$ are excited for the same critical Hartmann number and also -- if supercritical -- with the same growth rates (Fig.~\ref{fig1}). Despite  their simultaneous existence they can be excited as singles with different initial conditions. However, if the initial conditions are as neutral as possible with respect to a preferred helicity, in the majority of the cases one of the modes dominates after our experiences.

This behavior may have dramatic consequences with respect to the dynamo theory. Both the modes with $m=\pm 1$ have opposite helicity with the same total amount. The mode with $m=-1$ is identical to the mode with $m=1$ but in a left-hand system. The helicity of $m=1$ in the right-hand system equals the helicity of $m=-1$ in the left-hand system. So it is obvious that in one and the same coordinate system the sum of the helicity of $m=-1$ and $m=1$ is zero. As a consequence the instability of a toroidal field can only develop helicity if by some reasons one of the modes with $m=\pm 1$ dominates the other. There are the two possibilities that i) one mode dominates the other by chance (so as the matter dominates the antimatter) or ii) the existence of a poloidal field prefers one of the modes. We have shown that indeed in stellar radiation zones -- if the background field has a positive current helicity $\bar{\vec{B}} \cdot {\rm curl} \bar{\vec{B}}$ -- the resulting kinetic helicity $\langle \vec{u}' \cdot {\rm curl} \vec{u}'\rangle$ of the fluctuations is always negative (Gellert et al. 2011). A positive current helicity of the background field results if an axial field and an axial electrical current are parallel. A negative current helicity of the background field results if an axial field and an axial electric current are antiparallel. Hence, the resulting kinetic helicities  on the basis of current-driven instabilities have, therefore, opposite signs in opposite hemispheres of the model.

\section{Cylindric geometry}
We are interested in the  stability of a background field
$\bar{\vec{B}}= (0, B_\phi(R), B_0)$ with $B_0=\rm const$, and the flow
$\bar{\vec{u}}= (0,R\Omega(R), 0)$ in a dissipative conducting fluid rotating between two rigid cylinders.
 $\nu$ is  the kinematic viscosity and $\eta$ is 
the magnetic diffusivity, their ratio is the magnetic Prandtl number
${\rm Pm} =\nu/\eta$.
The  stationary background rotation law is 
$
\Omega=a +b/{R^2}$
with $a$ and  $b$ as constants.   $\Omega_{\rm{in}}$ and $\Omega_{\rm{out}}$ are the rotation
rates of the cylinders  and $B_{\rm{in}}$ and $B_{\rm{out}}$ are the azimuthal magnetic fields there.

One finds for the current helicity of the background field
$
\bar{\vec{B}}\cdot{\rm curl}\, \bar{\vec{B}}  \simeq A B_0,
$
which may be either positive or negative (and of course vanishes for the
current-free case $A=0$).  

The inner value $B_{\rm in}$ may be normalized with the uniform vertical field,
i.e. $
\beta ={B_{\rm in}}/{B_0}.
$
For a profile with  $B_{\rm{in}} =B_{\rm{out}}$ we have
$
 \bar{\vec{B}} \cdot {\rm curl}\, \bar{\vec{B}}\propto 1/\beta
$
for the normalized current helicity of the background field. The sign of $\beta$  determines the sign of the current helicity.  
As usual, the toroidal field amplitude is measured by the  Hartmann number ${\rm Ha}= B_{\rm in} D/ \sqrt{\mu_0 \, \rho \, \nu \, \eta}$  and the global rotation by the Reynolds number
$ {\rm Re}=\Omega_{\rm in}  D^2/\nu$
with $D=R_{\rm out} - R_{\rm in}$.  The Alfv\'en frequency  is 
$
\Omega_{\rm A} =B_{\rm in}/\sqrt{\mu_0\rho} D$.

The boundary conditions associated with the perturbation equations are no-slip
for the flows
and perfectly conducting for the fields.
 For all computations it is  $R_{\rm out}=2 R_{\rm in}$.
 
\begin{figure}[htb]
\begin{center}
 \mbox{\includegraphics[width=5cm]{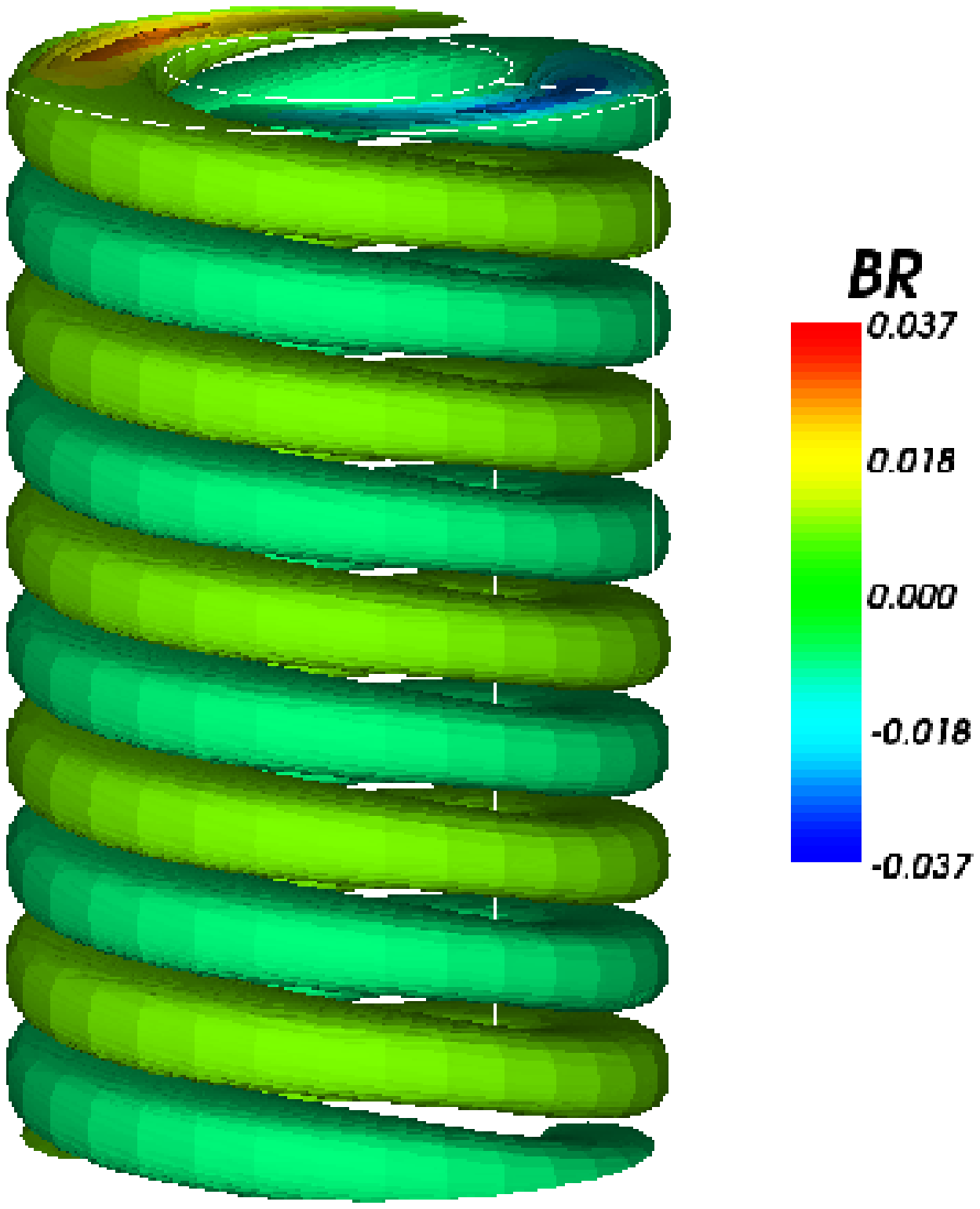}
  \includegraphics[width=5cm]{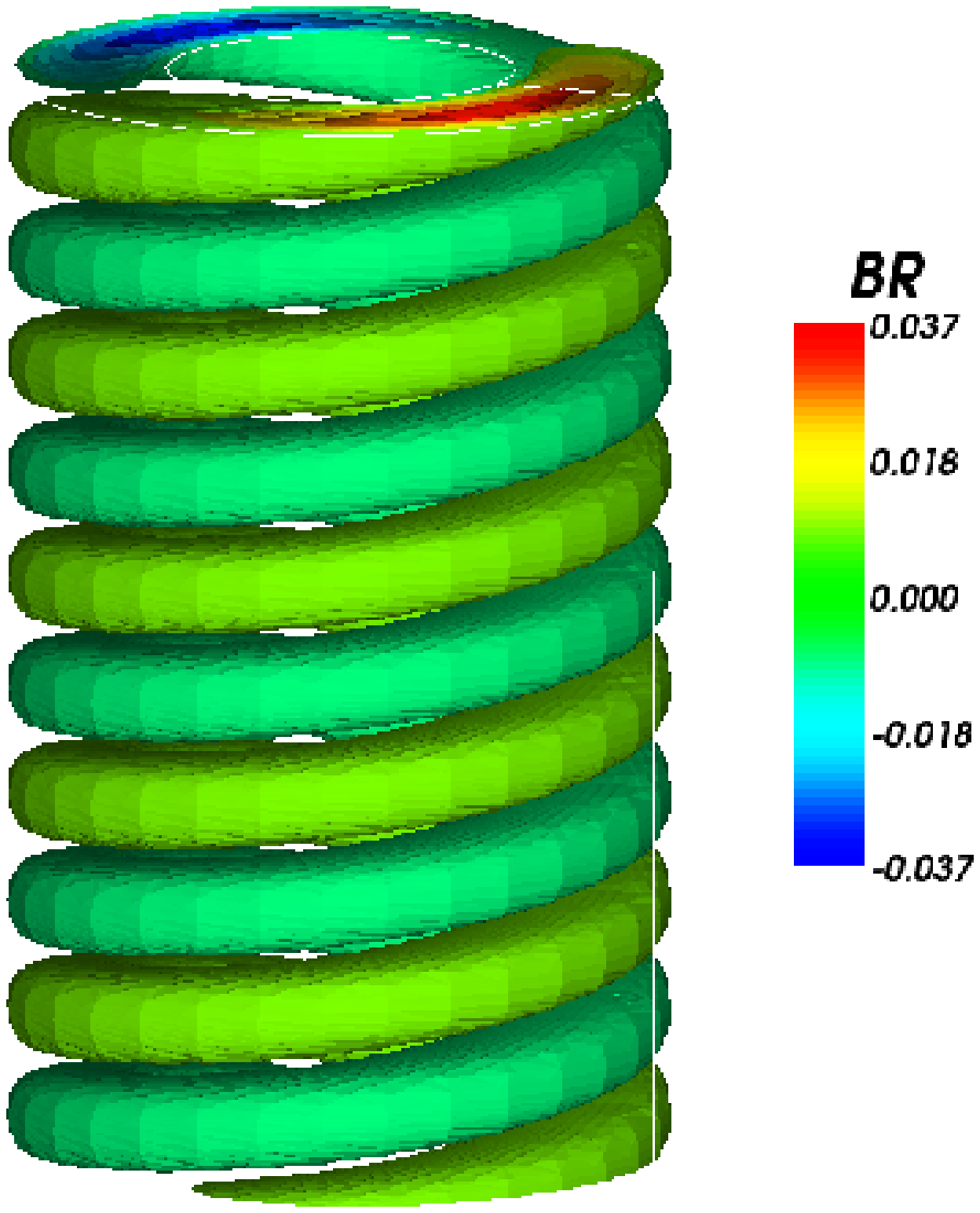}}
\end{center} 
\caption{\label{fig1} The two modes with opposite helicity values of the same amount which can be excited with different ini\-tial conditions.  ${\rm Re}=0$, ${\rm Ha}=200$,  $\beta=0$, ${\rm Pm}=1$ (from Gellert et al. 2011).}
\end{figure}

For purely toroidal fields the expected net helicities vanish as two nonaxisymmetric modes with $m=\pm 1$ exist with identical excitation conditions and the same amount of helicity but of opposite sign 
   (Fig. \ref{fig1}). 
  
\begin{figure}[htb]
\begin{center}
 \includegraphics[width=8cm]{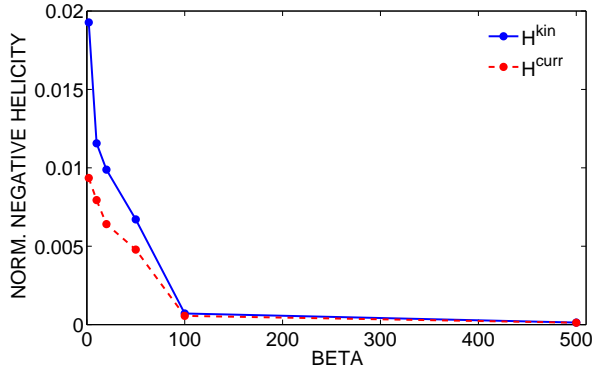}
\end{center} 
\caption{\label{fig2} The normalized negative kinetic helicity  and the negative current helicity  for the non\-axi\-symmetric perturbations as functions of  $\beta$. 
${\rm Re}=200$, ${\rm Ha}=100$, $\Omega_{\rm in}=2 \Omega_{\rm out}$,   ${\rm Pm}=1$.}
\end{figure}

We now compute both the helicities for helical fields with various pitch values $\beta$ which is inverse to the current helicity of the large-scale field. 
The astrophysically relevant case is $\Omega>\Omega_{\rm A}$ (which is not the classical realization of the Tayler instability).  The instability only exists for  nonrigid rotation so that we always work with a rotation law $\Omega \propto 1/R$. The radial profile of the toroidal field  has been modelled by the above mentioned most simple  profile with almost uniform toroidal field. Figure \ref{fig2}  gives the results (in units of $\Omega^2_{\rm A} D$). Both the kinetic  helicity $\langle \vec{u}' \cdot {\rm curl} \vec{u}'\rangle$ as well as the current helicity $\langle \vec{B}' \cdot {\rm curl} \vec{B}'\rangle$ are negative. They are decreasing functions of  $\beta$.

\begin{figure}[htb]
\begin{center}
 \mbox{
 \includegraphics[width=6.5cm,height=6.4cm]{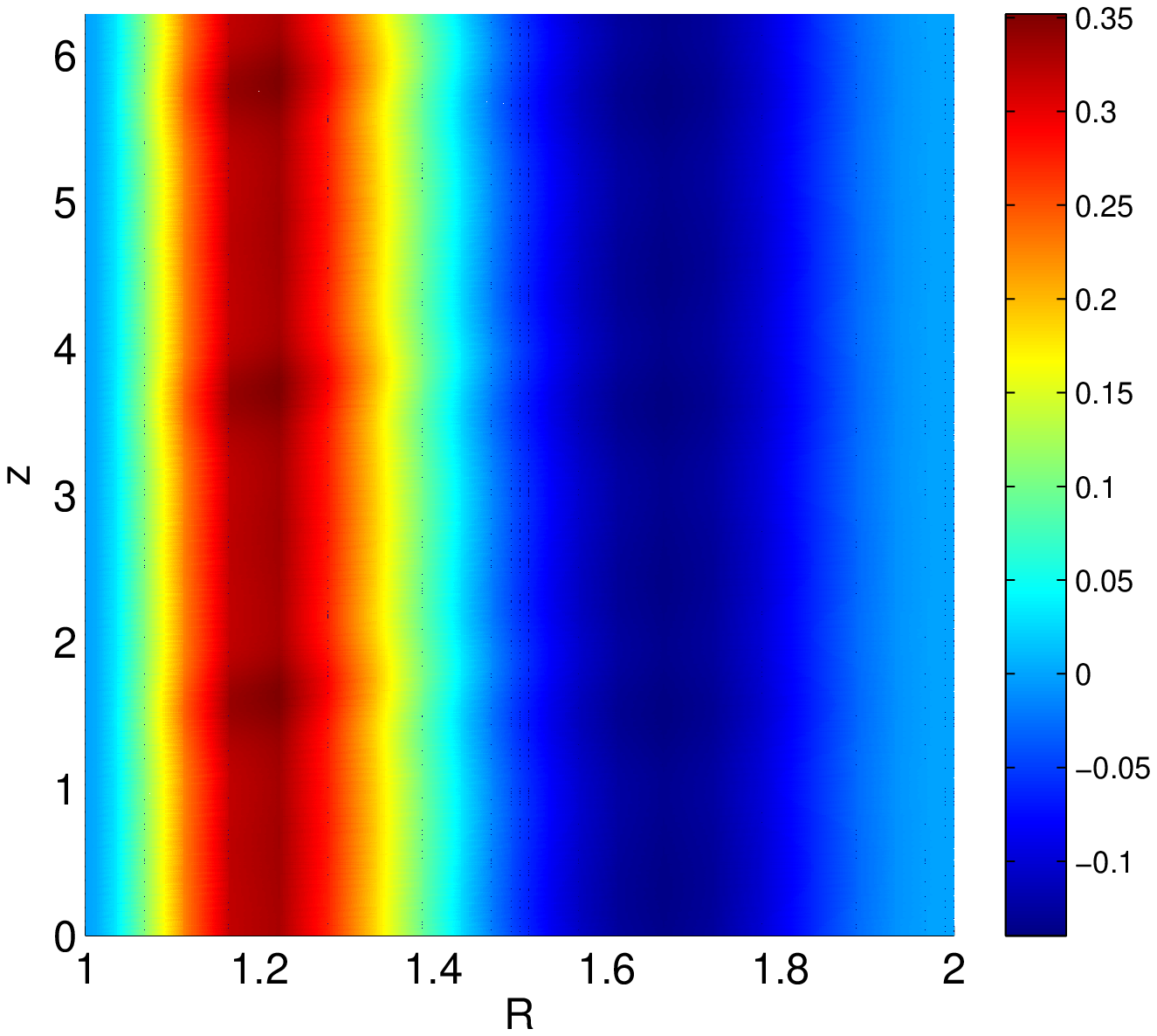}
  \includegraphics[width=6.5cm, height=6cm]{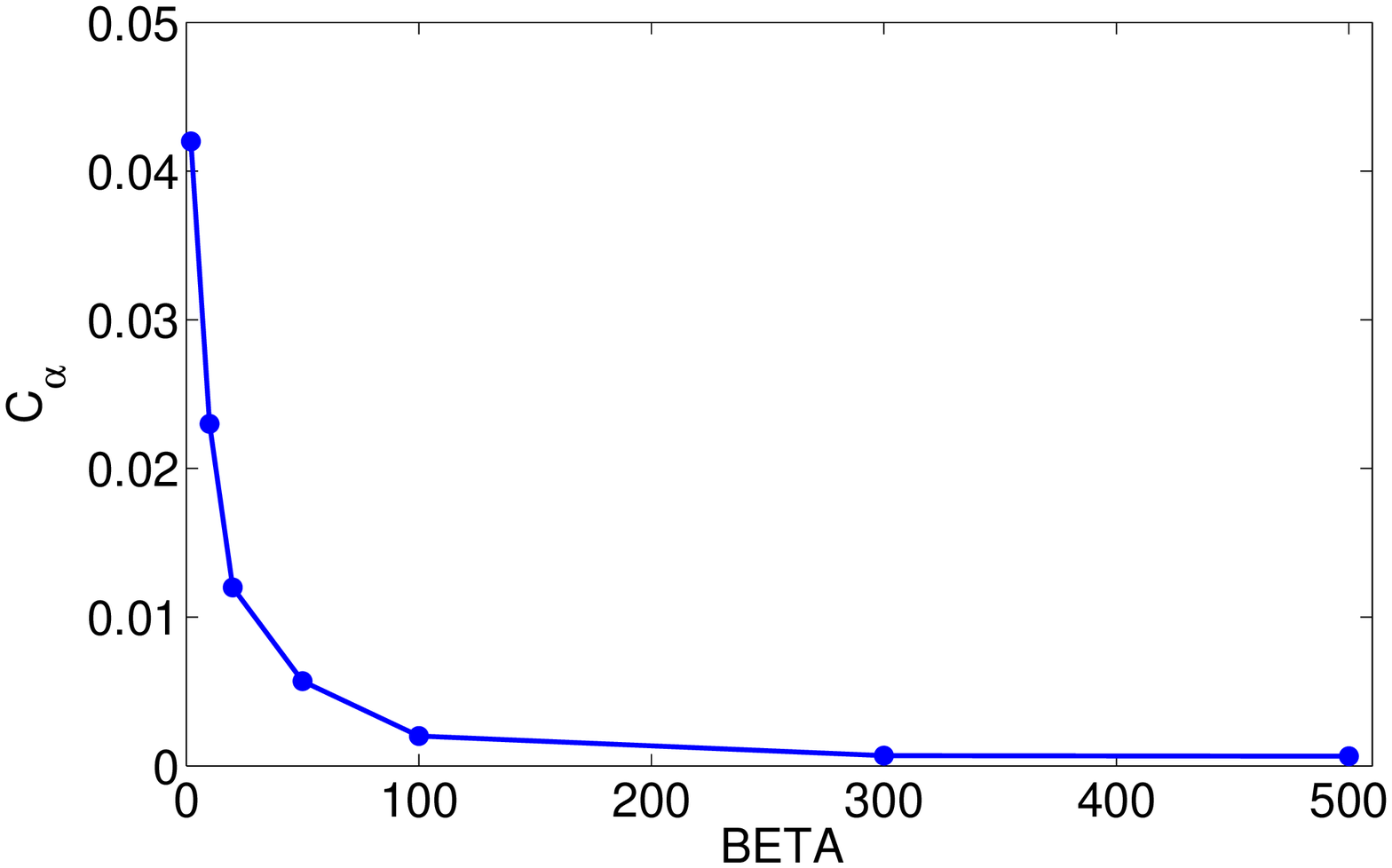}  }
\end{center} 
\caption{The  $\alpha$-effect  for $\Omega>\Omega_{\rm A}$. Left: $\beta=3$.     Right: The  dimensionless dynamo number $C_\alpha$ of the azimuthal $\alpha$-effect which decays  decays like $C/\beta$ with $C\simeq 0.05$.   ${\rm Pm}=1$. }
   \label{fig3}
\end{figure}

We have also calculated the $\alpha$-effect via the determination of the electromotive force $\langle \vec{u}'\times \vec{B}'\rangle$ by the fluctuations.  According to the  rule that the azimuthal $\alpha$-effect is anti-correlated with the (kinetic) helicity we expect the azimuthal $\alpha$-effect as positive for $\beta>0$.  
This is indeed the case (see Fig. \ref{fig3}, left). One finds positive and negative signs in the container but the positive values dominate so that in the average the azimuthal $\alpha$-effect is indeed positive. On the other hand, for negative $\beta$ we expect  positive small-scale kinetic helicity and negative $\alpha$-effect.

The example presented in Fig. \ref{fig3} refers to the pitch $\beta=3$. For higher values, i.e. smaller helicity of the background field, the resulting $\alpha$-effect becomes smaller and smaller. Figure \ref{fig3} (right) shows that the dynamo number $C_\alpha=\alpha_{\phi\phi}D/\eta$  runs as $C_\alpha\sim C/\beta$  with $C$ of order 10$^{-2}$. It is easy to show that this value is too small by two orders of magnitudes to allow the operation of a classical $\alpha\Omega$-dynamo (Gellert et al. 2011).

\section{Spherical geometry}
We start with  the amplification of fossil magnetic fields by shear and the magnetic-field  back-reaction. The two-dimensional, non-linear simulations thus start 
with an initial differential rotation and a purely poloidal 
magnetic field in a radiative stellar zone. 
The early phase of the simulation shows a generation and steep 
amplification of toroidal magnetic field through differential 
rotation. The generated  Lorentz force  redistributes the  angular momentum. This is why at the
same time of field amplification, the differential rotation 
starts to decrease, and the toroidal-field growth is thus limited.

\begin{figure}[htb]
\begin{center}
 \includegraphics[width=6.8cm]{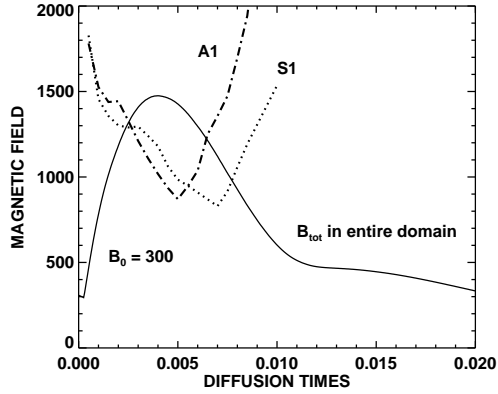}
\end{center} 
\caption{\label{fig4} The stability limits  of each snapshot of the axisymmetric evolution of a sufficiently strong initial field. The solid line gives the evolution of the toroidal field.
The stability limit is given for the nonaxisymmetric  perturbation patterns S1 and A1 (from Arlt \& R\"udiger 2011).}
  \end{figure}

Figure~\ref{fig4} shows the maximum magnetic field amplitude
in the spherical shell as a function of time. The 
Reynolds number is ${\rm Re} = 20\,000$ and the magnetic 
Prandtl number is ${\rm Pm}=1$.
The dotted and dot-dashed lines in Fig.~\ref{fig4} are the results
of a linear stability analysis of only the toroidal field.  By
S1 we refer to a velocity perturbation  which is symmetric 
with respect to the equator and has $m=1$, A1 is the corresponding 
antisymmetric perturbation. If at any given time both stability lines 
are above the solid line, the corresponding snapshot is stable against 
$m=1$ perturbations.  The stability lines cross the solid one at about 
$t=0.0023$ (see Arlt \& R\"udiger 2011). 

 The system is now perturbed in a 3D, nonlinear simulation at a 
 somewhat later time ($t_0=0.003$). The perturbation is applied to the magnetic field and has an 
azimuthal wave number of $m=1$ (and is symmetric with respect to the 
equator). The resulting flow is thus
antisymmetric and can be compared with the A1-mode in the linear
stability diagram in Fig.~\ref{fig4}.
\begin{figure}[b]
\begin{center}
 \mbox
 { \includegraphics[width=5.4cm,height=5.6cm
 ]{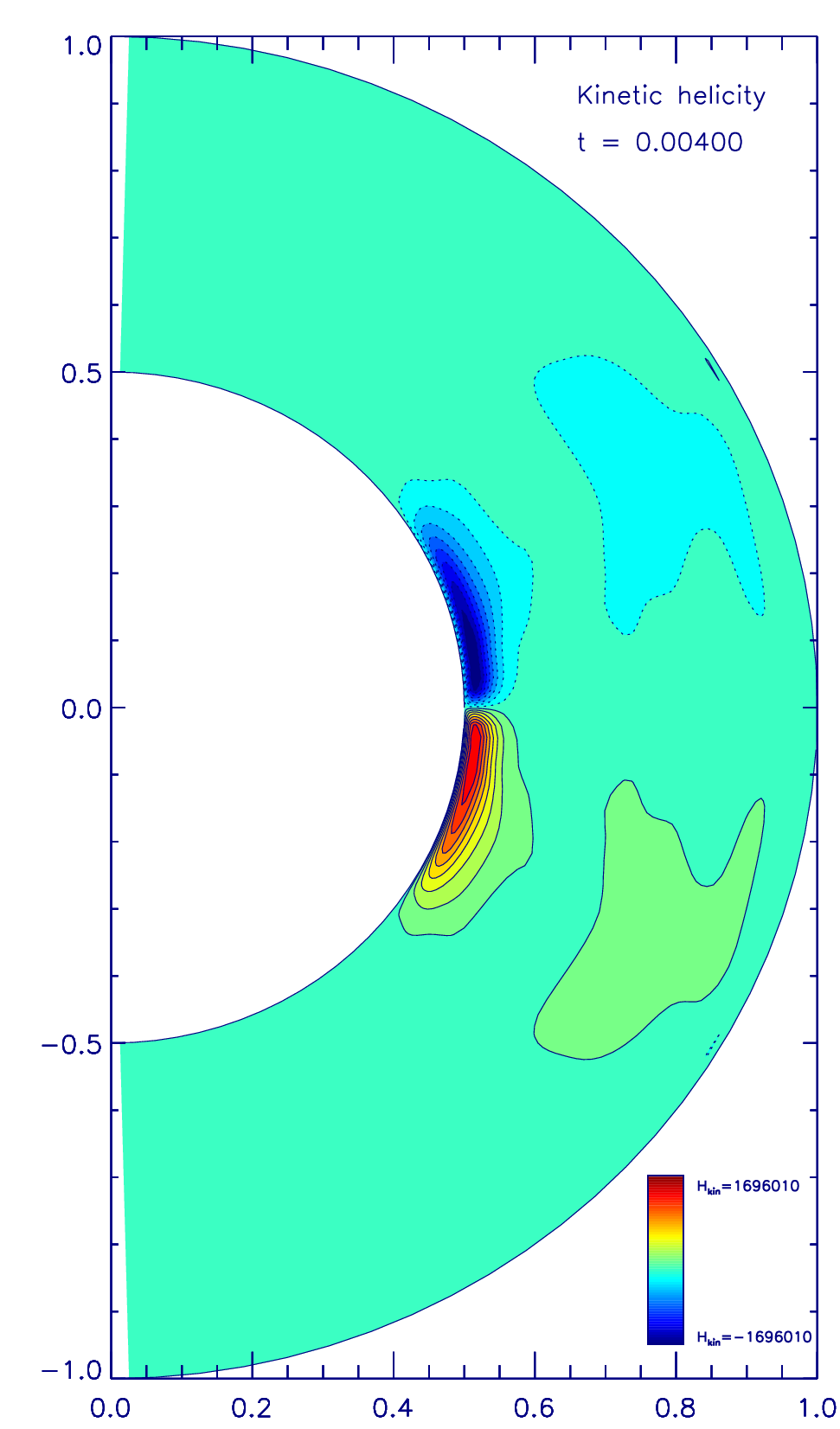}
  \includegraphics[width=5.4cm, height=5.6cm]{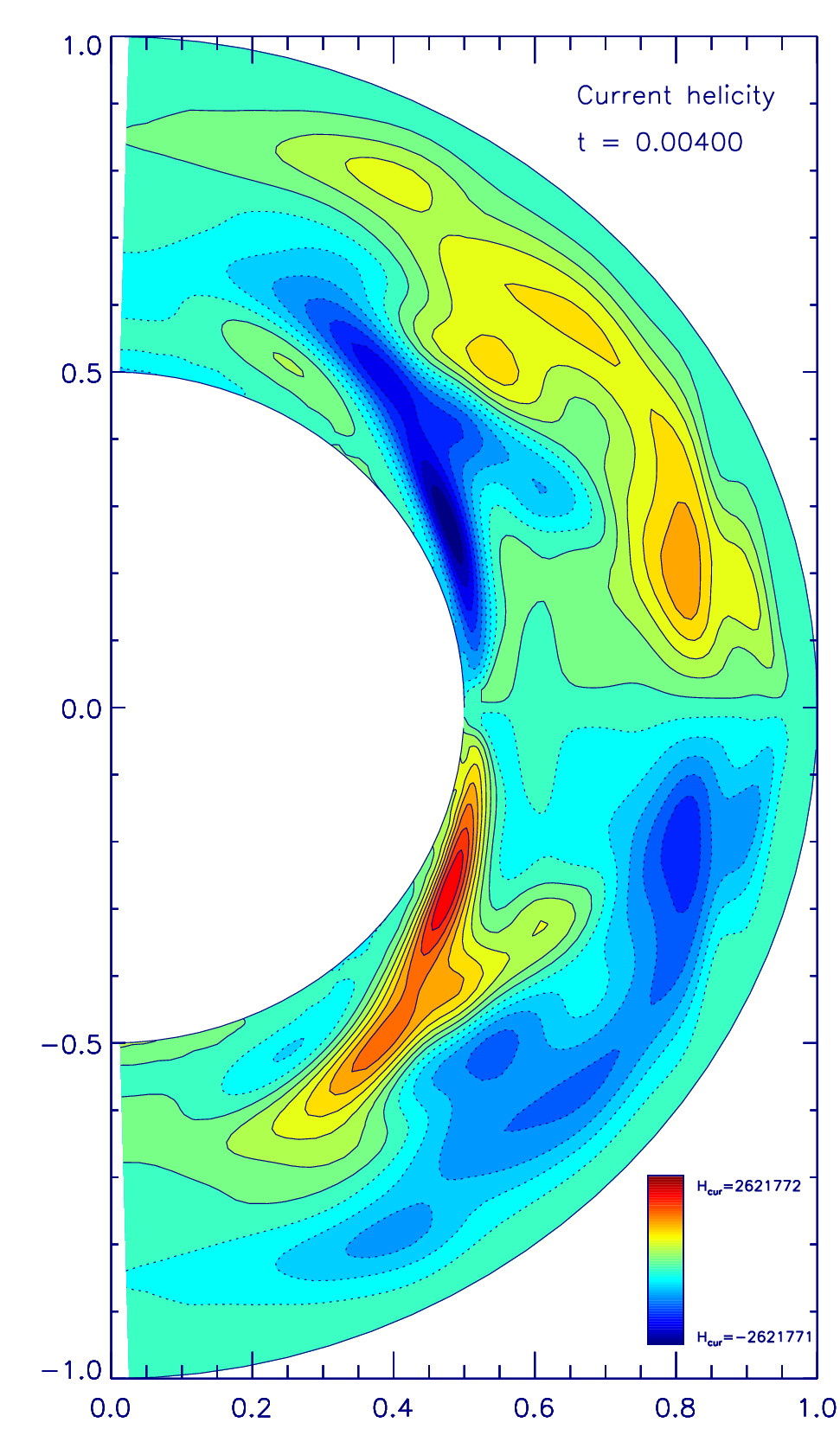}
  }
\end{center}
 \caption{ Left: Kinetic helicity, right: Current helicity. Both  after averaging over the azimuth. ${\rm Re} = 20\,000$.
 }
   \label{fig5}
\end{figure}

The spatial distribution of the resulting  helicity  is shown in Fig.~\ref{fig5}. Note the  concentration of both the 
helicities near the inner boundary  where
the tangent cylinder touches the inner sphere. We  conclude that a considerable part of the kinetic helicity  
in the system is due to  the presence 
of an inner sphere. 
This is slightly different for the current helicity. A considerable amount
of {\em positive\/} current helicity is measured in the bulk of the
northern hemisphere. The situation is unchanged in runs with 
perfect conductor boundary conditions at the inner radius.  Interestingly, Reshetnyak (2006) with convection simulations also finds negative
kinetic helicity along the tangent cylinder.  We assume that the negative helicity
near the tangent cylinder of the northern hemisphere is an
inner-boundary effect, neither related to convective nor Tayler
instability turbulence.
\begin{figure}[hb]
\begin{center}
\includegraphics[width=10.8cm,height=7cm]{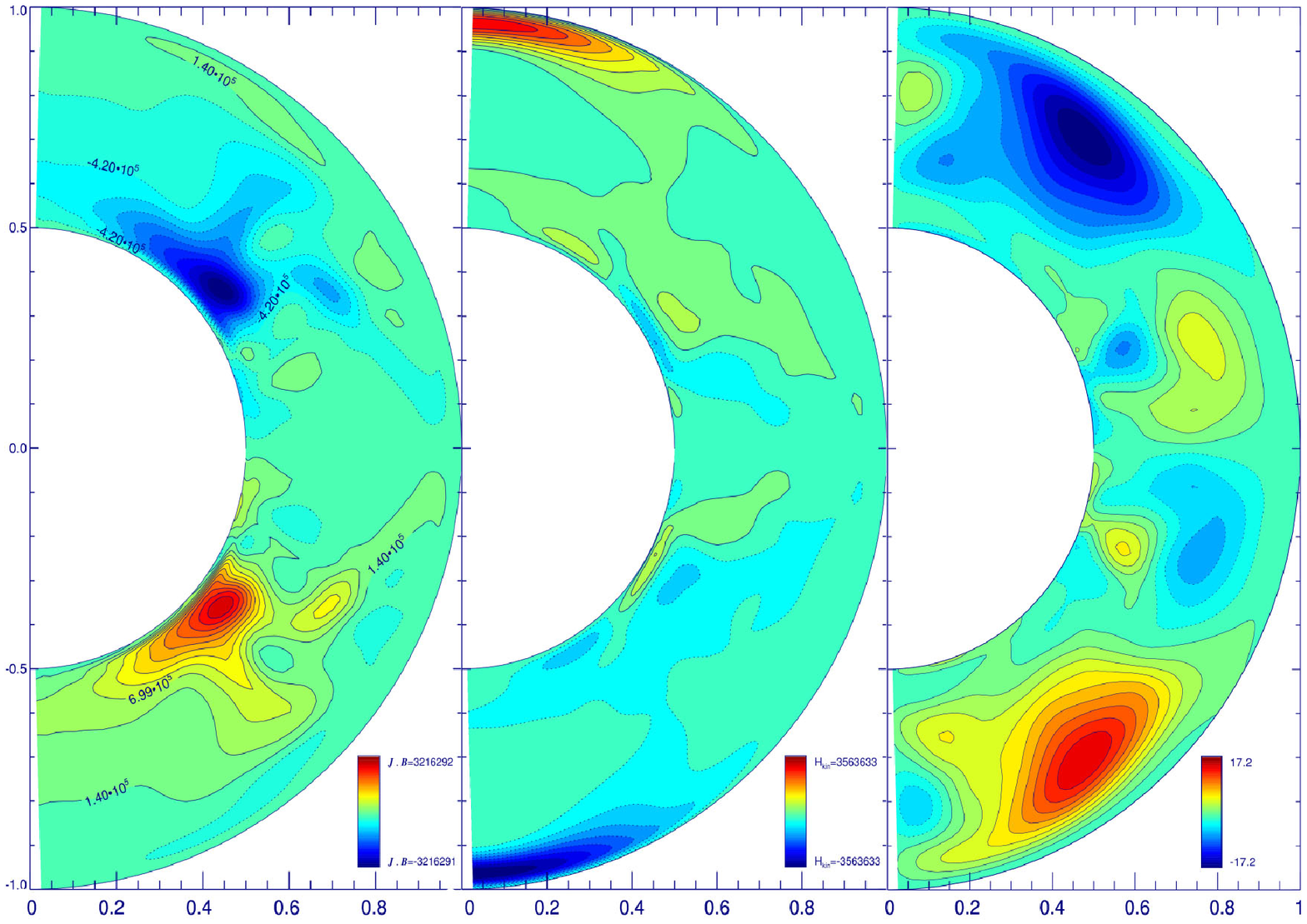}
\end{center}
 \caption{Left: $\bar{\vec{B}}\cdot {\rm curl} \bar{\vec{B}}$. Middle: kinetic helicity $\langle\vec{u}'\cdot {\rm curl} \vec{u}'\rangle$. Right: the azimuthal com\-ponent of the $\alpha$-effect. ${\rm Re}=0$. 
 }
   \label{fig6}
\end{figure}

Figure \ref{fig6} summarizes the consequences of a numerical experiment. In the unstable domain of Fig. \ref{fig4} the instability pattern is calculated without global rotation, i.e. for $\rm Re=0$. In this case the only existing pseudo-scalar is  $\bar{\vec{B}}\cdot {\rm curl} \bar{\vec{B}}$ like in the above theory in cylindric geometry. The question is whether we can find the same  relations between large-scale current helicity (left), kinetic helicity (middle) and $\alpha$-effect (right). Note first that indeed  the dominant role of the tangent cylinder for the  kinetic helicity and the $\alpha$-effect vanishes so that it is obvious that it is a rotation-induced boundary layer effect.  In the southern hemisphere the pseudo-scalar $\bar{\vec{B}}\cdot {\rm curl} \bar{\vec{B}}$ is positive while the kinetic helicity is negative (but concentrated to the pole). The corresponding $\alpha$-effect given in the right panel (and computed also by means of the test-field method) proves to be positive (see Arlt \& R\"udiger 2011). As it must, at the northern hemisphere all signs are opposite. Hence,  the model fulfills the same sign rules as in the above-discussed cylindric setup.

\section{Summary}
Unstable toroidal fields alone are not able to produce helicity and $\alpha$-effect.
It has been shown, however, that helicity and $\alpha$-effect are produced by unstable magnetic large-scale field patterns  which themselves possess current helicity  
$\bar{\vec{B}}\cdot {\rm curl} \bar{\vec{B}}$. This is insofar understandable as helicity and $\alpha$-effect are both pseudo-scalars which can only be nonvanishing  if in the global setup a nonvanishing large-scale pseudo-scalar  like $\bar{\vec{B}}\cdot {\rm curl} \bar{\vec{B}}$ exists. We want to stress, however, that the current helicity of the background field  is not the only possible pseudo-scalar  existing in magnetized stellar radiation zones  on which other forms of helicity and $\alpha$-effect may base.



\begin{thebibliography}{}

\bibitem[Arlt \& R\"udiger (2011)]{arru11}
Arlt, R., \& R\"udiger, G. 2011, \textit{MNRAS}, in press
\bibitem[Brott et al. (2008)]{br08}
Brott, I., Hunter, I., Anders, P., \& Langer, N. 2008, \textit{ AIPC}, 990, 273

\bibitem[Eggenberger et al. (2005)]{egg05}
Eggenberger, P., Maeder, A., \& Meynet, G. 2005, \textit{A\&A}, 440, L9

\bibitem[Gellert et al. (2011)]{ge11}
Gellert, M., R\"udiger, G., \& Hollerbach, R. 2011, \textit{Phys. Rev. E}, in  prep.

\bibitem[Reshetnyak (2006)]{resh06}
Reshetnyak, M.Yu. 2006, \textit{Physics of the Solid Earth}, 42, 449

\bibitem[R\"udiger \& Kitchatinov (1996)]{rueki96}
{R\"udiger, G., \& Kitchatinov, L.L.} 1996,
\textit{ApJ}, 466, 1078 

\bibitem[R\"udiger \& Schultz (2010)]{rueshu10}
{R\"udiger, G., \& Schultz, M.} 2010, 
\textit{Astron. Nachr.}, 331, 121

\bibitem[R\"udiger et al. (2009)]{rueetal09}
{R\"udiger, G., Gellert, M., \& Schultz, M.} 2009, 
\textit{MNRAS}, 399, 996


\bibitem[Spruit (2002)]{sp02}
Spruit, H.C. 2002, \textit{A\&A}, 381, 923



\bibitem[Yoon et al. (2006)]{yoon06}
Yoon, S.-C., Langer, N., \& Norman, C. 2006, \textit{A\&A}, 460, 199

\end{thebibliography}
\end{document}